\documentstyle[aps,twocolumn]{revtex}
\begin{document}

\title{Non-classical rotational inertia in the supersolid state}  

\draft
\author{Shun-ichiro Koh}
\address{ Physics Division, Faculty of Education, Kochi University  \\
        Akebono-cho, 2-5-1, Kochi, 780, Japan
        \thanks{e-mail address: koh@cc.kochi-u.ac.jp }
        }
\date{\today}

\maketitle
\draft

\maketitle
\begin{abstract}

An abrupt drop in the moment of inertia recently found in solid helium 4 
is explained in terms of dynamics of zero-point vacancies (ZPV). 
Mechanical decoupling of ZPV from the motion of the container due to  Bose 
statistics is developed to a macroscopic phenomenon by 
repulsive interaction. It gives a negative answer to the question whether 
BEC is a necessary condition for non-classical 
rotational inertia in a bulk three-dimensional system.
\end{abstract}
\pacs{PACS numbers: 67.80.-s, 67.40.-w, 05.30.-d}

 
Solid helium 4 has been termed a quantum crystal. On introduction of defects,  
it displays several anomalies not observed in classical crystals. 
Vacancies can move about unusually rapidly. 
Hence,  it is natural to expect that the quantum-mechanical fluctuation 
delocalizes vacancies at a sufficiently low temperature, which is called `zero-point 
vacancies' (ZPV)  \cite{and}. In such a `supersolid state', non-classical 
rotational properties are expected, which was 
argued with emphasis on the connectivity of the wave function \cite{leg1}. 
A number of experiments searching for anomalies in 
thermodynamical or mechanical properties have been performed, but ended with
null  results \cite{mei}.
Recently, an abrupt drop in the moment of inertia was found in the 
torsional oscillation measurements on solid helium 4 confined in a porous 
media \cite{cha} and on a bulk solid helium 4 \cite{kim}.  This discovery raises the question 
why other measurements  until now have found null results, and leads 
us to reconsider the definition of superfluidity and that of solids. 

 A natural way to discuss  superfluidity in a confined system is to focus on its 
rotational properties.  Anomalous behaviors of the system are normally attributed to the 
Bose condensate; in other words, off-diagonal long-range order 
(ODLRO). The question whether the Bose condensate is a necessary condition for 
superfluidity has been discussed in terms of academic interest. For
solid helium 4, however, this question is not an academic but a practical one. 
ZPV is still a hypothetical object, but is highly probable if solid helium 
4 shows superfluidity  \cite{mei}. 
 A fundamental feature of crystals lies in their periodicity in density; 
that is, diagonal long-range order (DLRO). One has to face a serious question whether 
 crystals remain stable while showing superfluidity that violates their periodicity. 
 
 In this paper, we assume the existence of ZPV obeying Bose statistics, 
 but do not assume its Bose condensate  \cite{sta}. Since vacancies are defects on the crystal 
 lattice, it is unlikely that crystals with a macroscopic number of 
 defects remain stable \cite{vie}.  Rather, in light of the definition of 
 solids, a more natural idea is that the number of ZPV is 
 large, but not macroscopically large in a solid helium 4. Hence, a 
 direction to be explored is the  
 possibility that under the influence of Bose statistics a  
 microscopic number of ZPV's create the non-classical rotational inertia without ODLRO. 
  For the dynamical properties such as superfluidity, 
  {\it the repulsive interaction generally enhances singular properties 
  due to Bose statistics. \/}  When the  container slowly rotates at low 
  temperature, some  microscopic regions emerge in solids, in which ZPV and 
 surrounding atoms decouple from the motion of container, thus
remaining at rest.  The  repulsive ZPV's are likely to spread 
uniformly in coordinate space. This feature makes ZPV outside of these  
regions behave similarly with ZPV at rest. Otherwise, the density of ZPV 
would become locally high, thus raising the interaction energy. 
  {\it The mechanical behavior of microscopic fractions of solids due to Bose 
 statistics (decoupling) is developed to a macroscopic phenomenon by the repulsive interaction. \/} 
In this paper, we stress that the non-classical rotational inertia (NCRI) 
is possible even in the case of no condensate by such a mechanism, and 
propose its formalism \cite{koh}.

 \underline{{\it Zero-point vacancies.\/} }
 Vacancies are usually regarded as localized objects which occasionally 
move from one position to another in crystals. It takes an energy 
$\epsilon _1$ for an atom  to diffuse to the surface. Further, the 
lattice distortion due to the vacancy raises the  energy by $\epsilon 
_2$. In quantum crystals, however, the delocalization of the  vacancy due 
to the quantum-mechanical fluctuation lowers the energy by $\Delta$. 
Hence, it takes $\epsilon _0=\epsilon _1+\epsilon _2-\Delta$ to make a 
vacancy in the crystal.  

In quantum crystals, there is a case in which $\Delta$ 
slightly exceeds $\epsilon _1+\epsilon _2$, thus leading to $\epsilon _0<0$. The 
possible states of such a vacancy are classified by the  quasi momentum 
$p$. With the tight-binding model applied to ZPV, its energy spectrum has a 
form as $\epsilon(p)=\epsilon _0+p^2/2m^*$, where $m^*$ is an effective 
mass of the vacancy involving surrounding atoms in its motion. 

When one vacancy approaches another in crystals, a lattice-distortion pattern by 
one vacancy is normally not consistent with that by another vacancy. (An  atom can 
not follow two displacement patterns that contradict each other.) This situation 
results in an effective repulsive interaction $U$ between the two 
vacancies. 

Let us consider a crystal with ZPV as $ H= H_0+H_{Z}$
where $H_0$ represents a hamiltonian of a crystal with no vacancies, 
$H_Z$ a hamiltonian of ZPV
\begin{equation}
 H_{Z}=\sum_{p}\epsilon (p)\Phi_{p}^{\dagger}\Phi_{p}
   +U\sum_{p,p'}\sum_{q}\Phi_{p-q}^{\dagger}\Phi_{p'+q}^{\dagger}\Phi_{p'}\Phi_p , еее
	\label{е}
\end{equation}е
and $\Phi$ denotes an annihilation operator of ZPV as a spinless boson 
\cite{mod}. The motion of vacancies is expressed by Green's function as 
$G(\omega,p)=1/(\omega-\epsilon (p))$.  Multiple scattering between vacancies
 is expressed by t-matrix as $t=U/(1+UG)$.  In the tight-binding 
picture of ZPV, the band width of ZPV has an order of  $\Delta$, thus 
$G\simeq \Delta ^{-1}$. As $U\rightarrow \infty$, $t$ approaches 
$1/G\simeq \Delta$. Hence, an energy $\hat{\epsilon}$ of the repulsive ZPV takes a 
form such as $\hat{\epsilon}=-|\epsilon _0|+n\Delta$,  where $n$ is a 
number density of ZPV. The increase in the number of ZPV stops when  
$\hat{\epsilon}$ reaches zero, that is, $n=|\epsilon _0|/\Delta$, 
following which the number of ZPV may change only through their 
fusion to the surface of crystal. Its relaxation time is much 
longer than the time required to establish equilibrium for a given number 
of ZPV. At low temperature, ZPV behaves as a repulsive Bose 
quasiparticle in the equilibrium state.

For the stability of solids, even in the case of ZPV ($\epsilon _0<0$) 
$\Delta$ only slightly exceeds $\epsilon _1+\epsilon _2$,  which 
would otherwise lead to the collapse of a solid. Because of 
$|\epsilon _0|=|(\epsilon _1+\epsilon _2)-\Delta|\ll 
\Delta$, the number density satisfies $n=|\epsilon _0|/\Delta\ll 1$. 
Hence, it is unlikely that the number of ZPV is a macroscopic one. 

  \underline{{\it Moment of inertia.\/}}
 By the torsional oscillator experiment, one measures the moment of inertia  $I$ of 
 a sample as a response to the restoring force of the torsion rod. As long 
 as the external force is small,  $I$ is determined by the  rotational 
 property of the system without that force. 
  
 Consider  a solid helium 4 in a uniform rotation round   z-axis  with 
 angular velocity $\Omega$.  A hamiltonian in a coordinate system rotating 
  with the body is $H-\mbox{\boldmath $\Omega$}\cdot \mbox{\boldmath $L$}$, where  
$\mbox{\boldmath $L$}$ is the total angular momentum. 
The rotation is equivalent to the application of the probe 
acting on the sample. The perturbation $-\mbox{\boldmath $\Omega$}\cdot \mbox{\boldmath $L$}$ 
is cast in the form $-\sum_{i} (\mbox{\boldmath $\Omega$}\times 
\mbox{\boldmath $$r})\cdot \mbox{\boldmath $p$}е$ , where 
$\mbox{\boldmath $\Omega$}\times 
\mbox{\boldmath $r$} \equiv \mbox{\boldmath $v$}_d( \mbox{\boldmath $r$})$ is the drift velocity of the 
sample at point $\mbox{\boldmath $r$}$. With the current density $ 
\mbox{\boldmath $J$}(\mbox{\boldmath $r$})$, the perturbation has a form as
\begin{equation}
-\mbox{\boldmath $\Omega$}\cdot \mbox{\boldmath $L$}
   =-\int  \mbox{\boldmath $v$}_d( \mbox{\boldmath $r$})\cdot 
   \mbox{\boldmath $J$}(\mbox{\boldmath $r$}) d\mbox{\boldmath $r$}.е
\end{equation}е
Because of $ div\mbox{\boldmath $v$}_d( \mbox{\boldmath $r$})=0 $, 
Eq.(2) says that the rotation represented by  $\mbox{\boldmath $v$}_d( \mbox{\boldmath $r$})$
 acts as a transverse-vector 
probe to the excitations in solids \cite{noz}. We need the linear 
response of the system $\mbox{\boldmath $J$}(\mbox{\boldmath $r$})$ to 
$\mbox{\boldmath $v$}_d( \mbox{\boldmath $r$})$. 
 Since a solid normally rotates like a rigid body, we can use the mass density  
 $\rho$ in $\mbox{\boldmath $J$} (\mbox{\boldmath $r$})=\rho\mbox{\boldmath $v$}_d( \mbox{\boldmath $r$})$. 
 Microscopically, however,  one must begin with the 
transverse susceptibility $ \chi ^T (q,\omega)$ of the system. 
(By definition, the mass density  $\rho$ is regarded as the longitudinal 
response to a force as $\rho=nm=\chi^L(0,0)$.) One assumes  
the spatial homogeneity of the  sample during rotation. Hence, $\bf{J}(\bf{r})$ is approximated as
\begin{equation}
     \mbox{\boldmath $J$} (\mbox{\boldmath $r$})=\left[\lim_{q\rightarrow 0}
    \chi ^T (q,0)\right] \mbox{\boldmath $v$}_d( \mbox{\boldmath $r$} )е  
\end{equation}е
Using Eq,(3) and $ \mbox{\boldmath $v$}_d=(-\Omega y, \Omega x,0) $ in the right-hand 
side of Eq.(2), and $ \mbox{\boldmath $\Omega$} =(0,0,\Omega) $ in its left-hand side, 
one obtains the angular momentum $L_z$, hence the moment of inertia $I_z=L_z/\Omega$ as
\begin{equation}
     I_z=  \chi^T(0,0)\int_{V}(x^2+y^2) d\bf{r} .е
\end{equation}е

(A) The susceptibility of a normal solid satisfies $\chi^T(0,0)=\chi 
^L(0,0)$. Hence, replacing $\chi^T(0,0)$ by $\rho$ in Eq.(4),
one obtains the definition of classical $I_z^{c}$. 
 
  The generalized susceptibility is defined as
\begin{equation}
	\chi_{\mu\nu}(q,\omega )=\frac{q_{\mu}q_{\nu}}{q^2е}\chi^L(q,\omega)     
	                    +\left(\delta_{\mu\nu}-\frac{q_{\mu}q_{\nu}}{q^2е}\right)е\chi^T(q,\omega) .
	\label{е}
\end{equation}е

 For the later use, we define a  term proportional to $q_{\mu}q_{\nu}$ in 
 $\chi_{\mu\nu}$ as $\hat{\chi}_{\mu\nu}$ so that
 \begin{eqnarray}
       	\chi_{\mu\nu}(q,\omega)&=&\delta_{\mu\nu}\chi^T(q,\omega)
	                  +q_{\mu}q_{\nu}\left(\frac{\chi^L(q,\omega)-\chi^T(q,\omega)}{q^2е}\right)е \nonumber \\ 
	                          &\equiv& \delta_{\mu\nu}\chi^T+\hat{\chi}_{\mu\nu}(q,\omega), 
	\label{е}
\end{eqnarray}е
 and rewrite Eq.(4) as 
\begin{equation}
     I_z= I_z^{c} - \lim_{q\rightarrow 0} 
                      \left[\frac{q^2}{q_{\mu}q_{\nu}е}\hat{\chi}_{\mu\nu}(q,0)\right]е
                         \int_{V}(x^2+y^2) d\bf{r} .е
\end{equation}е

(B) At the onset of ZPV, according to  $H=H_0+ H_{Z}$, 
 the susceptibility $\chi(q,\omega)$ splits into $\chi_0(q,\omega)$ and  
$\chi^{ZPV}(q,\omega)$ \cite{app}.  The normal part gives  
 no contribution to NCRI in Eq.(7) due to $\chi_0^L(0,0)=\chi_0^T(0,0)$, 
 whereas the validity of $\chi^{ZPV,L}(0,0)=\chi ^{ZPV,T} (0,0)$ must be examined.
For the appearance of NCRI, the balance between the longitudinal 
and the transverse excitation must be destroyed on a macroscopic scale 
($q\rightarrow 0$). On this point, one can trace back to Feynman's 
physical argument on how the Bose-statistical coherence suppresses the 
transverse excitation, and destroys this balance \cite{fey}. In his 
argument, not the Bose condensate but Bose statistics is essential. 
 
(C) To see the effect of the repulsive interaction, let us consider $\chi_{\mu\nu}$ of 
 ZPV as an ideal Bose gas.  Within the linear response, it is defined as
\begin{eqnarray}
	 \lefteqn{\chi^{ZPV}_{\mu\nu}(q,\omega _n) } \nonumber\\
	            && 	=\frac{1}{Vе}\int_{0}^{\betaе}d\tau \exp(i\omega_n\tau)
	                          <S|T_{\tau}J_{\mu}(q,\tau)J_{\nu}(q,0)|S>еее,
\end{eqnarray}е 
where $J_{\mu}(q,\tau)$ is a current of ZPV ($\hbar =1$) \cite{cur} 
\begin{equation}
	J_{\mu}(q,\tau)=\sum_{p,n} 
	\left(p+\frac{q}{2е}\right)_{\mu}\Phi_p^{\dagger}\Phi_{p+q}e^{i\omega _n\tau}еее,
	\label{е}
\end{equation}е
and $|S>$ is a ground sate of  $H_0+\sum_{p}\epsilon (p)\Phi_{p}^{\dagger}\Phi_{p}$.
 Its $\hat{\chi}_{\mu\nu}(q,\omega)$ is given by 
\begin{eqnarray}
	 \lefteqn{\hat{\chi}^{ZPV}_{\mu\nu}(q,\omega) } \nonumber\\
	            && =-\frac{q_{\mu}q_{\nu}}{4е}е
	                          \sum_{p}\frac{f(\epsilon (p))-f(\epsilon (p+q))}
	                                       {\omega+\epsilon (p)-\epsilon (p+q)е}е,
\end{eqnarray}е 
where $f(\epsilon (p))$ is the Bose-Einstein distribution.

(1) If ZPV would form the condensate, 
 $f(\epsilon (p))$ in Eq.(10) is a macroscopic 
number for $p=0$ and nearly zero for $p\ne 0$. Thus, in the sum over $p$ in the 
right-hand side of Eq.(10), only two terms corresponding to $p=0$ and 
$p=-q$ remain, with a result that 
\begin{equation}
	\hat{\chi}^{ZPV}_{\mu\nu}(q,0)=m^*n_0е\frac{q_{\mu}q_{\nu}}{q^2е}е,
\end{equation}е 
where $n_0$ is the number density of the condensate. 
 Equation.(7) with Eq.(11) shows NCRI. (The relative degree  
 $(I_z-I_z^{c})/I_z^{c}$ of the deviation from the classical moment of inertia
 would be simply proportional to the number density of the condensate.)

(2) When ZPV forms no condensate, however, the sum over $p$ in 
Eq.(10) is carried out by replacing it with an integral, and one notices 
that $q^{-2}$ dependence disappears in the result. 
Hence, without the repulsive interaction, BEC is a necessary condition of NCRI.

\underline{{\it Non-classical rotational inertia.\/} }
 We will formulate a mechanism of NCMI under the more realistic condition for a solid helium 4.
  Instead of $|S>$ in Eq.(8), a ground state  $|G>$ of $H_0+H_Z$  
 must be used as 
  \begin{eqnarray}
       \lefteqn{ <G|T_{\tau}J_{\mu}(x,\tau)J_{\nu}(0,0)|G> } \nonumber\\ 
       && =  \frac{<S|T_{\tau}\hat{J}_{\mu}(x,\tau)\hat{J}_{\nu}(0,0)
 	              exp\left[-i\int_{0}^{1/\betaе}d\tau H_I(\tau)е\right]|S>е}
 	        {<S|exp\left[-i\int_{0}^{1/\betaе}d\tau H_I(\tau)е\right]|S>е}е,
	\label{е}
\end{eqnarray}е
 where $H_I(\tau)$ represents the repulsive interaction.
 Due to the repulsive interaction $U$ contained in $exp(-i\smallint  H_I(\tau)d\tau)$, 
 the scattering between ZPV's frequently occurs  as illustrated by an upper 
 bubble with a dotted line in Fig.1(a). The current-current response 
 tensor $J_{\mu}(x,\tau)J_{\nu}(0,0)$ is depicted by a lower bubble. 
 
Since $|G>$ is a ground state under the influence of Bose statistics,
 the perturbation must be developed 
in such a way that as the  order of the expansion increases, the  
susceptibility gradually includes a new effect due to Bose statistics.
  An important feature is that ZPV in $J_{\mu}(x,\tau)J_{\nu}(0,0)$ 
  and ZPV in the upper bubble due to the repulsive interaction form a coherent wave function as a whole. 
 Hence, we must seriously  consider the influence of Bose statistics on 
the graph like Fig.1(a). When one of the two  ZPV's in the lower bubble 
and that in the upper bubble have the same momentum 
($p=p'$), and the other ZPV in the two bubbles have the  
 same momentum ($p+q=p'+q'$), a graph made by 
interchanging these two types of ZPV's must be included in the expansion of 
$\chi^{ZPV}_{\mu\nu}$. The interchange of ZPV lines with $p$ and $p'$($=p$) in Fig.1(a) yields 
Fig.1(b). Further, the interchange of ZPV lines  with 
$p+q$ and $p+q'$($=p+q$) yields Fig.1(c), a graph linked by the 
repulsive interaction, whose contribution to $\chi^{ZPV}_{\mu\nu}$ is given by 
\begin{eqnarray}
	 \lefteqn{	\chi^{ZPV,(1)}_{\mu\nu}(q,\omega)} \nonumber\\
	            && =U\sum_{p}(p+\frac{q}{2е}е)_{\mu}(p+\frac{q}{2е}е)_{\nu}
	                 \left[-\frac{f(\epsilon (p))-f(\epsilon (p+q))}
	                            {\omega+\epsilon (p)-\epsilon (p+q)е} \right]^2еее.
\end{eqnarray}е
With decreasing temperature, the coherent wave function grows to a large 
size, and the interchange of ZPV due to Bose statistics like Fig.1 occurs 
many times. Hence, one can not ignore the higher-order term $\chi_{\mu\nu}^{ZPV,(n)}$ 
which is significant in the larger coherent wave function.  Among many ZPV's with various momentums,  
ZPV's at rest with $p=0$ play a key role, leading to the following form  
\begin{equation}
	\hat{\chi}_{\mu\nu}^{ZPV,(n)}(q,0)=\frac{q_{\mu}q_{\nu}}{4е}еU^nF_{\beta}(q)^{n+1},
	\label{е}
\end{equation}е
where
\begin{equation}
	F_{\beta}(q)=  \frac{(\exp(-\beta\mu)-1)^{-1}-(\exp(\beta [\epsilon (q)-\mu])-1)^{-1}}
	                         {\epsilon (q)ее} е,
	\label{е}
\end{equation}е
and  $\mu$  is a chemical potential.
$F_{\beta}(q)$ is a positive monotonously decreasing function of $q^2$ which approaches zero as 
$q^2\rightarrow \infty$. As $\mu\rightarrow 0$, $F_{\beta}(q)$ increases at any $q$.
 An expansion form of $F_{\beta}(q)$ around $q^2=0$  is given by
\begin{eqnarray}
	  F_{\beta}(q) &=&\frac{\beta\exp(\beta\mu)}{(1-\exp(\beta\mu))^2е}  \nonumber \\
	               &-& \frac{\beta ^2}{2е}
	                 \exp(\beta\mu)\frac{(1+\exp(\beta\mu))}{(1-\exp(\beta\mu))^3е}\epsilon(q)+\cdots.е
\end{eqnarray}е

When $q^2\hat{\chi}_{\mu\nu}/q_{\mu}q_{\nu}$ has a finite value at 
$q\rightarrow 0$ in Eq.(7), that is, when a power series in $U$ 
\begin{equation}
	\hat{\chi}^{ZPV}_{\mu\nu}(q,0)=\frac{q_{\mu}q_{\nu}}{4е}е\sum_{n=0}^{\inftyе}U^nF_{\beta}(q)^{n+1},
	\label{е}
\end{equation}е
 diverges as $q^{-2}$ at $q\rightarrow 0$, the moment of inertia shows 
 a non-classical behavior. In normal solids ($\beta\mu \ll 0$), a small 
 $F_{\beta}(q)$ guarantees the convergence of  $\hat{\chi}^{ZPV}_{\mu\nu}(q,0)$,  
  with a result that,
\begin{equation}
	\hat{\chi}^{ZPV}_{\mu\nu}(q,0)=\frac{q_{\mu}q_{\nu}}{4е}е
	                 \frac{F_{\beta}(q)}{1-UF_{\beta}(q)е}.ее
	\label{е}
\end{equation}е
With decreasing temperature ($\mu\rightarrow 0$), however, a gradual increase  
of  $F_{\beta}(q)$ makes the higher-order term significant in Eq.(17), finally 
leading to the divergence  of power series in $\hat{\chi}^{ZPV}_{\mu\nu}(q,0)$. 
 The convergence condition is first violated at $q=0$ when 	
$UF_{\beta}(0)=1$, that is,  
\begin{equation}
     U\beta=4\sinh ^2\left(\frac{\beta\mu}{2е}е\right)е .
	\label{е}
\end{equation}е
 From now, we call $T_0$ satisfying Eq.(19) an onset temperature of  
NCRI.  Equation (19) implies that {\it a solid with ZPV shows NCRI 
without the condensate\/} ($\mu\ne 0$). At $T=T_0$, substituting 
 Eq.(16) in Eq.(18) with the aid of Eq.(19), we find as a leading term
\begin{equation}
	\hat{\chi}^{ZPV}_{\mu\nu}(q,0)=\frac{m^*}{2\sinh |\beta _0\mu |е}е
	                              \frac{q_{\mu}q_{\nu}}{q^2е}е.
\end{equation}е
In view of Eq.(6), Eq.(20) leads to 
$\chi^{ZPV,L}(q,0)\ne\chi ^{ZPV,T}(q,0)$ at $q\rightarrow 0$. 
 Since the large-scale interchange of bosons plays  an essential role for deriving 
 Eq.(20), this result confirms Feynman's intuitive argument on the 
 influence of Bose statistics on the excitations. 
Using Eq.(20) in Eq.(7) with the aid of Eq.(19), we obtain 
\begin{equation}
     I_z= I_z^{c} - \frac{m^*}{\sqrt{ \frac{U}{k_BT_0е}е \left(4+\frac{U}{k_BT_0е}ее\right)ее}е}
                         \int_{V}(x^2+y^2) d\bf{r} ,е
\end{equation}е
which implies that $(I_z-I_z^{c})/I_z^{c}$  for superfluidity 
 by ZPV does not allow the one-body interpretation as $\rho _s/\rho$, but reflects the 
 many-body effect due to Bose statistics and the repulsive interaction. 
Since ZPV is not a particle in the literal sense, but represents the motion of 
surrounding atoms, this is a natural result.

 \underline{ {\it Discussion. \/}}
Superfluidity is a complex of phenomena, and therefore  
 has some different definitions such as, (1) persistent 
current without friction, (2) the Hess-Fairbank effect, (3) quantized 
circulation, (4) almost no friction on moving objects in the system below the critical velocity, 
(5) peculiar collective excitations and (6) the Josephson 
effect \cite{leg2}. (Strictly speaking, the experiment by Kim and Chan 
adds a new definition of superfluidity to the list.) 
Since these phenomena (1) $\sim$ (6) are observed to occur together in most materials 
exhibiting superfluidity (including superconductors), it is difficult to elucidate the exact 
nature of the relationships between these definitions. From the standpoint  
 of this paper, there is no reason to expect that all superfluid-like behaviors 
corresponding to these definitions should occur in quantum crystals. We 
must therefore  classify them into two categories. The  first is the   
phenomenon which an interplay of Bose statistics and the repulsive force creates 
without the condensate, and the second is that for which the condensate 
is necessary. The first and the second corresponds to an observable and an 
unobservable phenomenon in quantum crystals, respectively.  (From this 
viewpoint, we must put a new interpretation on some `null results' in 
earlier measurements of mechanical properties of a solid helium 4 \cite{the}.)
We can regard a solid helium 4, together with a helium 4 film \cite{rep}, as 
 appropriate material for examining the relationship between the 
different definitions of superfluidity.

╩


е


%
%
 \begin{figure}
 \caption{ The first-order Feynman diagram of the current-current 
           response tensor.  The black and the white small circle  
           represents a vector and a scalar vertex, respectively. 
           The dotted line denotes the repulsive interaction.
       }
 \label{Fig.1}
 \end{figure}

%
%


\begin{thebibliography}{99}
\bibitem{sta} Although the Bose condensate is a most peculiar 
           feature of Bose statistics, more fundamental is a fact that 
           particles obey Bose statistics even when BEC is absent.
\bibitem{mod} If the defects are dislocations, the model must be 
       improved to deal with the extended objects.
\bibitem{app} It is assumed that the motion of vacancies can be 
         separately considered from other excitations in crystals.
\bibitem{cur} Since  $\mbox{\boldmath $v$}_d$ is the neutral 
       counterpart of a vector potential $A_{\mu}$, 
       Eq.(9) has the same structure as that in the charged Bose gas.
\bibitem{the} Since the condensate is absent, there is little probability of 
        detecting signs in thermodynamical properties such as the temperature dependence of pressure.
\end{thebibliography}
\end{document}